\documentstyle[twoside,fleqn,espcrc2]{article}

\newcommand{\AmS}{{\protect\the\textfont2
  A\kern-.1667em\lower.5ex\hbox{M}\kern-.125emS}}

\title{Neutrino Masses and Grand Unification}

\author{G. Altarelli \address{Theoretical Physics Division, CERN, \\ 
 1211 Geneva 23, Switzerland \\ 
 and\\
Universit\`a di Roma Tre, Rome, Italy \\ }%
	}
       
\begin{document}

\begin{abstract}
We discuss some models of neutrino masses and mixings in the context
 of fermion masses in Grand Unified Theories.
\end{abstract}

\maketitle

\def\beq{\begin{equation}} 
\def\eeq{\end{equation}}
 \def\bea{\begin{eqnarray}} 
 \def\eea{\end{eqnarray}}
\def\bq{\begin{quote}} 
\def\eq{\end{quote}}

\def\AJ{{\it Astrophys.J.} } 
\def\AJL{{\it Ap.J.Lett.} } 
\def\AJS{{\it Ap.J.Supp.} } 
\def\AM{{\it Ann.Math.} } 
\def\AP{{\it Ann.Phys.} } 
\def\APJ{{\it Ap.J.} } 
\def\APP{{\it Acta Phys.Pol.} }
\def\ASAS{{\it Astron. and Astrophys.} } 
\def\BAMS{{\it Bull.Am.Math.Soc.} } 
\def\CMJ{{\it Czech.Math.J.} } 
\def\CMP{{\it Commun.Math.Phys.} } 
\def\FP{{\it Fortschr.Physik} } 
\def\HPA{{\it Helv.Phys.Acta} } 
\def\IJMP{{\it Int.J.Mod.Phys.} } 
\def\JMM{{\it J.Math.Mech.} } 
\def\JP{{\it J.Phys.} } 
\def\JCP{{\it J.Chem.Phys.} } 
\def\LNC{{\it Lett. Nuovo Cimento} } 
\def\SNC{{\it Suppl. Nuovo Cimento} } 
\def\MPL{{\it Mod.Phys.Lett.} } 
\def\NAT{{\it Nature} } 
\def\NC{{\it Nuovo Cimento} }
\def\NP{{\it Nucl.Phys.} } 
\def\PL{{\it Phys.Lett.} } 
\def\PR{{\it Phys.Rev.} } 
\def\PRL{{\it Phys.Rev.Lett.} } 
\def\PRTS{{\it Physics Reports} } 
\def\PS{{\it Physica Scripta} } 
\def\PTP{{\it Progr.Theor.Phys.} } 
\def\RMPA{{\it Rev.Math.Pure Appl.} } 
\def\RNC{{\it Rivista del Nuovo Cimento} }
\def\SJPN{{\it Soviet J.Part.Nucl.} } 
\def\SP{{\it Soviet.Phys.} } 
\def\TMF{{\it Teor.Mat.Fiz.} }
\def\TMP{{\it Theor.Math.Phys.} } 
\def\YF{{\it Yadernaya Fizika} } 
\def\ZETF{{\it Zh.Eksp.Teor.Fiz.} }
\def\ZP{{\it Z.Phys.} } 
\def\ZMP{{\it Z.Math.Phys.} }

\parskip 0.3cm

\def\gappeq{\mathrel{\rlap {\raise.5ex\hbox{$>$}} {\lower.5ex\hbox{$\sim$}}}}

\def\lappeq{\mathrel{\rlap{\raise.5ex\hbox{$<$}} {\lower.5ex\hbox{$\sim$}}}}



\noindent





Recent data from Superkamiokande \cite{SK} have provided a more solid experimental basis for neutrino
oscillations as an explanation of the atmospheric neutrino anomaly \cite{Ronga}. In addition the solar neutrino deficit
\cite{bel}, observed by several experiments, is also probably an indication of a different sort of neutrino oscillations.
Results from the laboratory experiment by the LSND collaboration \cite{LSND} can be considered as a possible indication of
yet another type of neutrino oscillation.  Neutrino oscillations imply neutrino masses. The extreme smallness of neutrino
masses in comparison with quark and charged lepton masses indicate a different nature of neutrino masses, linked to
lepton number non conservation and the Majorana nature of neutrinos. Thus neutrino masses provide a window on the very large
energy scale where lepton number conservation is violated and on GUTs. The new experimental evidence on
neutrino masses could also give an important feedback on the problem of quark and charged lepton masses, as all these
masses are possibly related in GUTs. In particular the observation of a nearly maximal mixing angle for
atmospheric neutrinos is particularly significant. Perhaps also solar neutrinos may occur with
large mixing angle. At present solar neutrino mixings can be either large or very small, depending on which particular
solution will eventually be established by the data. Large mixings are very interesting because a first
guess was in favour of small mixings in the neutrino sector in analogy to what is observed for quarks. If confirmed, single
or double maximal mixings can provide an important hint on the mechanisms that generate neutrino masses  \cite{ram}.

The experimental status of neutrino oscillations is still very preliminary. While the evidence for the
existence of neutrino oscillations from solar and atmospheric neutrino data is rather convincing by now, the values of
the mass squared differences $\Delta m^2$ and mixing angles are not firmly established. For solar neutrinos, for example,
three or even four possible solutions are still possible \cite{fogli}\cite{hall}.  Another issue which is still open is the
claim by the LSND collaboration of an additional  signal of
neutrino oscillations in an accelerator experiment \cite{LSND}. This claim was not so-far supported by a second recent
experiment, Karmen \cite{Karmen}, but the issue is far from being closed. Given the present experimental uncertainties the
theorist has to make some assumptions on how the data will finally look like in the end. Here we tentatively assume
that the LSND evidence will disappear (for the alternative option, see, for example, refs.\cite{cald}). If so then we only 
have two oscillations frequencies, which can be given in terms of the three known species of light neutrinos without
additional sterile kinds (i.e. without weak interactions, so that they are not excluded by LEP). We then take for granted
that the frequency of atmospheric neutrino oscillations will remain well separated from the solar neutrino frequency, even
for the MSW large angle solution. We also assume that the electron neutrino does not participate in the atmospheric
oscillations, which (in absence of sterile neutrinos) are interpreted as nearly maximal
$\nu_{\mu}\rightarrow\nu_{\tau}$ oscillations as indicated by the Superkamiokande \cite{SK} and Chooz
\cite{Chooz} data. However the data do not exclude a non-vanishing $U_{e3}$ element. In the Superkamiokande allowed
region the bound by Chooz
\cite{Chooz} amounts to  $|U_{e3}|\lappeq 0.2$ \cite{fogli,hall}.

In summary, by now we have a substantial evidence that neutrinos are massive. From a strict minimal standard model point
of view neutrino masses could vanish if no right handed neutrinos existed (no Dirac mass) and lepton number was
conserved (no Majorana mass). In GUTs both these assumptions are violated. The right handed neutrino is required in all
unifying groups larger than SU(5). In SO(10) the 16 fermion fields in each family, including the right handed neutrino,
exactly fit into the 16 dimensional representation of this group. This is really telling us that there is something in
SO(10)! Thus SO(10) must at least appear as a classification group at $M_{Planck}$, if not as a symmetry group at $M_{GUT}$.
The breaking of
$|B-L|$, B and L conservation is also a generic feature of GUTs. In fact, the see-saw mechanism \cite{ssm} explains
the smallness of neutrino masses in terms of the large mass scale where $|B-L|$ and L conservation laws are violated. Thus,
neutrino masses are important as a probe into the physics at the GUT scale, as would be proton decay, although in a less
direct way. For example, heavy Majorana neutrinos could be part of the explanation of baryogenesis. If baryogenesis at the
weak scale is excluded by the data it can occur at or just below the GUT scale, after inflation. But only that part with
$|B-L|>0$ would survive and not be erased at the weak scale by instanton effects. Thus baryogenesis at $kT\sim
10^{12}-10^{15}~GeV$ needs B-L non conservation at some stage like for $m_\nu$ with Majorana neutrinos. The two
effects could be related if baryogenesis arises from leptogenesis via $\nu$ decay \cite{lg} then converted into baryogenesis
by instantons. Present results on neutrino masses are compatible with this picture \cite{buch}. Thus the possibility of
baryogenesis at a large energy scale has been boosted by the recent results on neutrinos. 

Oscillations only determine squared mass differences and not masses. The case of three nearly degenerate neutrinos
is the only one that could in principle accomodate neutrinos as hot dark matter together with solar and atmospheric
neutrino oscillations. For a cosmologically significant fraction of hot dark matter, the common mass should be around 1-3 eV.
The solar frequency could be given by a small 1-2 splitting, while the atmospheric frequency could be given by a still small
but much larger 1,2-3 splitting.  Note that we
are assuming only two frequencies, given by $\Delta_{sun}\propto m^2_2-m^2_1$ and
$\Delta_{atm}\propto m^2_3-m^2_{1,2}$. A strong constraint arises in the degenerate case from neutrinoless double beta
decay which requires that the ee entry of
$m_{\nu}$ must obey
$|(m_{\nu})_{11}|\leq 0.2-0.5~{\rm eV}$ \cite{dbeta}. As observed in ref. \cite{GG}, this bound can only be 
satisfied if
double maximal mixing is realized, i.e. if also solar neutrino oscillations occur with nearly maximal mixing.
Note that for degenerate masses with $m\sim 1-3~{\rm eV}$ we need a relative splitting $\Delta m/m\sim
\Delta m^2_{atm}/2m^2\sim 10^{-3}-10^{-4}$ and an even smaller one for solar neutrinos. It is not simple
to imagine a natural mechanism compatible with unification and the see-saw mechanism to arrange such a
precise near symmetry, stable under running down from the GUT to the weak scale \cite{ello},\cite{BHKR}.

If neutrino masses are smaller than for cosmological relevance, we can have the hierarchies $|m_3| >> |m_{2,1}|$
or $|m_1|\sim |m_2| >> |m_3|$. We prefer the first case, because for quarks and charged leptons one
mass eigenvalue, the third generation one, is largely dominant. Thus the dominance of $m_3$ for neutrinos
corresponds to what we observe for the other fermions.  In this case, $m_3$ is determined by the atmospheric
neutrino oscillation frequency to be around $m_3\sim0.05~eV$. By the see-saw mechanism $m_3$ is related to some
large mass M, by $m_3\sim m^2/M$. If we identify m with the Higgs vacuum expectation value or the top mass
then M turns out to be around $M\sim 10^{15}~GeV$, which is indeed consistent with the
connection with GUTs.

Here we concentrate on models
with three light neutrinos, large light neutrino mass splittings and large mixings \cite{us}. In general large
splittings correspond to small mixings because normally only close-by states are strongly mixed. The requirement of large
splitting and large mixings imposes a condition of a vanishing determinant. For example, in a 2 by 2 context, the matrix
\beq 
m\propto 
\left[\matrix{ x^2&x\cr x&1    } 
\right]~~~~~. 
\label{md0}
\eeq has eigenvalues 0 and $1+x^2$ and for $x$ of 0(1) the mixing is large. Thus in the limit of neglecting small mass
terms of order $m_{1,2}$ the demands of large atmospheric neutrino mixing and dominance of $m_3$ translate into the
condition that the 2 by 2 subdeterminant 23 of the 3 by 3 mixing matrix vanishes in some approximate limit. The problem is to
show that this vanishing can be arranged in a natural way without fine tuning \cite{us12}, \cite{all}.

One possible mechanism is based on asymmetric neutrino Dirac matrices, with  a large
left-handed mixing already present in the Dirac matrix. In ref.\cite{us3}, we argued that in a SU(5)
GUT left-handed mixings for leptons tend to correspond to right-handed mixings for d quarks (in a basis where u quarks are
diagonal). Since large right-handed mixings for quarks are not in contrast with experiment, viable GUT models can be
constructed following this mechanism that correctly reproduce the data on fermion masses and mixings \cite{achi}. 

If, for some reason, one prefers symmetric matrices (for example, one could want to
preserve left-right symmetry at the GUT scale) then assuming that $m_D$ is nearly diagonal in the basis where
charged leptons are diagonal, large mixings could arise from the Majorana sector (for example, by dominance of a light right
handed neutrino, with large couplings to $\nu_{mu}$ and $\nu_{tau}$\cite{bar},\cite{king}) . In a recent paper \cite{us4}, we
have presented examples where a nearly maximal mixing is created from almost nothing: all relevant matrices entering in the
see-saw mechanism are diagonal, yet the resulting mixing is large. Or large neutrino mixings could be generated by an
enhancement of formally small terms \cite{Lola}.  This is because a typical small term in quark or charged lepton mass
matrices is of the order of the Cabibbo angle
$\lambda\sim 0.22$ which is not that small.  

In conclusion the fact that some neutrino mixing angles are large, while surprising at the start, was eventually found to
be well be compatible, without any major change, with our picture of quark and lepton masses within GUTs. Rather it
provides us with new important clues that can become sharper when the experimental picture will be further clarified.

\end{document}